\documentclass[journal=jctc,manuscript=article]{achemso}
\usepackage[version=3]{mhchem} 

\usepackage{xr}
\makeatletter
\newcommand*{\addFileDependency}[1]{
  \typeout{(#1)}
  \@addtofilelist{#1}
  \IfFileExists{#1}{}{\typeout{No file #1.}}
}
\makeatother

\newcommand*{\myexternaldocument}[1]{
    \externaldocument{#1}
    \addFileDependency{#1.tex}
    \addFileDependency{#1.aux}
}

\myexternaldocument{SI}
\usepackage{amsmath}
\usepackage{amssymb}
\usepackage{graphicx,color}
\usepackage{url}
\usepackage{bbm}
\usepackage{algorithm}
\usepackage{algpseudocode}

\usepackage{multirow}

\newtheorem{thm}{Theorem}

\newcommand{\ii}{\mathbbm{i}}

\author{Xiaoxiao Xiao}
\author{Hewang Zhao}
\author{Jiajun Ren}
\author{Wei-Hai Fang}
\author{Zhendong Li}
\email{zhendongli@bnu.edu.cn}
\affiliation[University]
{Key Laboratory of Theoretical and Computational Photochemistry, Ministry of Education, College of Chemistry, Beijing Normal University, Beijing 100875, China}

\title{Physics-Constrained Hardware-Efficient Ansatz on Quantum Computers that is Universal, Systematically Improvable, and Size-consistent}

\begin{document}

\begin{tocentry}

\includegraphics{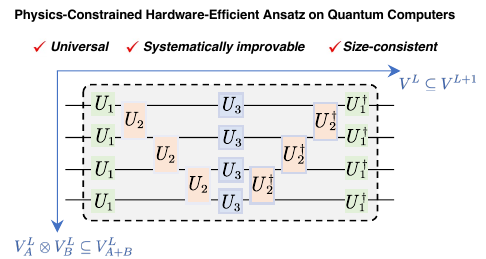}




\end{tocentry}


\begin{abstract}
Variational wavefunction ans\"{a}tze are at the heart of solving
quantum many-body problems in physics and chemistry.
Previous designs of hardware-efficient ansatz (HEA) 
on quantum computers are largely based on heuristics and
lack rigorous theoretical foundations.
In this work, we introduce a physics-constrained approach for designing HEA
with rigorous theoretical guarantees by imposing a few fundamental constraints.
Specifically, we require that the target HEA to be universal, systematically improvable, and size-consistent, which is an important concept in quantum many-body theories for scalability, but has been overlooked in previous designs of HEA.
We extend the notion of size-consistency to HEA, and
present a concrete realization of HEA that satisfies all these fundamental constraints while only requiring linear qubit connectivity. The developed physics-constrained HEA is superior to other heuristically designed HEA in terms of both accuracy and scalability, as demonstrated numerically for the Heisenberg model and some typical molecules. In particular, we find that restoring size-consistency can significantly reduce the number of layers needed to reach certain accuracy.
In contrast, the failure of other HEA to satisfy these constraints severely limits their scalability to larger systems with more than ten qubits.
Our work highlights the importance of incorporating physical constraints into the design of HEA for efficiently solving many-body problems on quantum computers.
\end{abstract}


\section{Introduction}
Efficient simulation of quantum many-body problems is an enduring frontier in computational physics and chemistry\cite{martin2016interacting}. 
Among many different approaches, the variational method represents a powerful and versatile
technique to tackle quantum many-body problems. A wealth of variational wavefunction
ans\"{a}tze on classical computers have been developed over the past decades. Prominent examples include Slater determinants, Gutzwiller wavefunction\cite{gutzwiller1963effect},
Jastrow wavefunction\cite{jastrow1955many}, tensor network states\cite{white1992density,verstraete2004renormalization,shi2006classical,vidal2008class,verstraete2008matrix,schollwock2011density,orus2014practical}, and neural network states\cite{carleo2017solving,choo2018symmetries,liang2018solving,choo2019two,hibat2020recurrent,sharir2020deep,barrett2022autoregressive}. Thanks to the rapid
progress on quantum hardware\cite{arute2019quantum,wu2021strong}, the variational quantum eigensolver (VQE)\cite{peruzzo2014variational,mcclean2016theory}, which is
a hybrid quantum-classical approach for solving quantum many-body problems\cite{tilly2022variational},
has attracted much attention\cite{cao2019quantum,mcardle2020quantum,bauer2020quantum,cerezo2021variational}. The central component of VQE is the preparation of a trial wavefunction on quantum computers, which
ultimately determines the accuracy of the variational computation.
Compared to the development of variational ans\"{a}tze
on classical computers, the exploration of wavefunction ans\"{a}tze
on quantum computers is still in its infancy.

Available variational ans\"{a}tze on quantum computers developed so far can be broadly classified into two categories: physics/chemistry-inspired ans\"{a}tze and hardware-efficient ans\"{a}tze (HEA), each with its own advantages and disadvantages. The chemistry-inspired unitary coupled cluster (UCC) ansatz\cite{anand2022quantum} is the first ansatz proposed for determining molecular ground states on quantum computers\cite{peruzzo2014variational}, which is motivated by the great success of the traditional coupled cluster theory on classical computers\cite{bartlett2007coupled}. However, it quickly becomes impractical for large molecules on the current noisy intermediate-scale quantum (NISQ) hardware\cite{preskill2018quantum}, since the circuit depth scales as $O(N^4)$ with respect to the number of qubits $N$\cite{whitfield2011simulation,seeley2012bravyi,hastings2015improving}.
Many efforts have been devoted to reduce the complexity of UCC, resulting in
several descendants of UCC such as the unitary paired CC ansatz\cite{lee2018generalized}, 
the unitary cluster Jastrow ansatz\cite{matsuzawa2020jastrow}, and some adaptive variants\cite{ryabinkin2018qubit,grimsley2019adaptive,tang2021qubit,yordanov2021qubit}.
Another type of physics-inspired ansatz is the Hamiltonian variational ansatz (HVA)\cite{wecker2014gate,wecker2015progress}, which is problem specific and widely used for model systems\cite{wiersema2020exploring}. 
When applied to general cases such as the molecular Hamiltonian in quantum chemistry, it suffers from the same problem as UCC.

Hardware efficient ans\"{a}tze were originally 
proposed as a more practical alternative on near-term quantum devices\cite{kandala2017hardware}. It takes the following form 
\begin{eqnarray}
|\Psi\rangle = U_L(\vec{\theta}_L)\cdots U_1(\vec{\theta}_1)|\Phi_0\rangle
\triangleq
\prod_{l=1}^{L} U_l(\vec{\theta}_l)|\Phi_0\rangle,\label{eq:HEA}
\end{eqnarray}
where $|\Phi_0\rangle$ is a reference state and the repeating unit
$U_l(\vec{\theta}_l)$ is a parameterized quantum circuit (PQC)
composed of gates that are native on quantum hardware\cite{kandala2017hardware},
such as single-qubit rotation gates and two-qubit entangling gates (see Fig. \ref{fig:ansatz}a).
Given the hardware constraints, there is still a great deal
of freedom in choosing the layout of the
circuit block $U_l(\vec{\theta}_l)$.
So far, the architectures of HEA have been designed mostly by heuristics, 
and there is no theoretical guarantee for their performance. 
Furthermore, the optimization of HEA is challenging\cite{bittel2021training} as the number of
qubits $N$ and the number of layers $L$ increases, due to
the proliferation of low-quality local minima\cite{anschuetz2022quantum} and the exponential
vanishing of gradients, known as the  'barren plateau' phenomenon\cite{mcclean2018barren}.
These problems severely limit the scalability of HEA beyond small systems.
Therefore, it is highly desirable to design a variational ansatz with rigorous theoretical basis,
while at the same time being hardware-efficient on near-term devices\cite{d2023challenges,motta2023bridging}.

In this work, we present a new way to design HEA with rigorous theoretical guarantees by imposing fundamental constraints. This is inspired by the remarkably successful way to 
design exchange-correlation (XC) functionals in density functional theories (DFT) 
by requiring the XC functionals to satisfy exact constraints\cite{kaplan2023predictive},
which has lead to reliable non-empirical XC functionals for a wide range of systems\cite{perdew1996generalized,sun2015strongly}.
We expect that designing HEA in a similar way can lead to
more systematic construction of variational ans\"{a}tze on quantum computers.
The remaining part of the paper is organized as follows:
First, we introduce some fundamental constraints for HEA.
Specifically, we require the ansatz to be universal, systematically improvable, and size-consistent (or multiplicatively separable), which is an important property for scalability in quantum many-body theory. Then, we present one concrete realization, which satisfies all these requirements while requiring 
only linear qubit connectivity with 
nearest neighbor interactions. Consequently, a layerwise optimization strategy is
introduced to take full advantage of the systematic improvability of
the proposed ansatz, which is shown to alleviate the barren plateau problem.
The effectiveness of this ansatz is demonstrated for
the Heisenberg model and some typical molecules.
The comparison with other HEA shows that
incorporating physical constraints into the
design of HEA is a promising way to design
more efficient and scalable variational ans\"{a}tze on quantum computers.

\section{Theory and algorithm}
\subsection{Fundamental constraints for HEA}
We first introduce some fundamental constraints that a good HEA should satisfy. Given a system $A$, suppose the variational space of HEA with $L$ layers in Eq. \eqref{eq:HEA} is denoted by $V_A^L$, we impose the following four basic constraints for the possible form of $U_l(\vec{\theta}_l)$:

{\bf (1) Universality}: any quantum state should be approximated arbitrarily well by the designed HEA with a sufficiently large number of layers $L$.

{\bf (2) Systematic improvability}: $V_A^L$ should be included in $V_A^{L+1}$ for any $L$, i.e. $V_A^L\subseteq V_A^{L+1}$. This guarantees that the variational space is systematically expanded, and the variational energy converges monotonically as $L$ increases, i.e. $E_{A}^{L+1}\le E_{A}^{L}$. A simple sufficient condition for the systematic improvability
is that there exists a set of parameters $\vec{\theta}_l$
such that $U_l(\vec{\theta}_l)=I$.
    
\begin{figure}[H]
\centering
\includegraphics[width=0.6\textwidth]{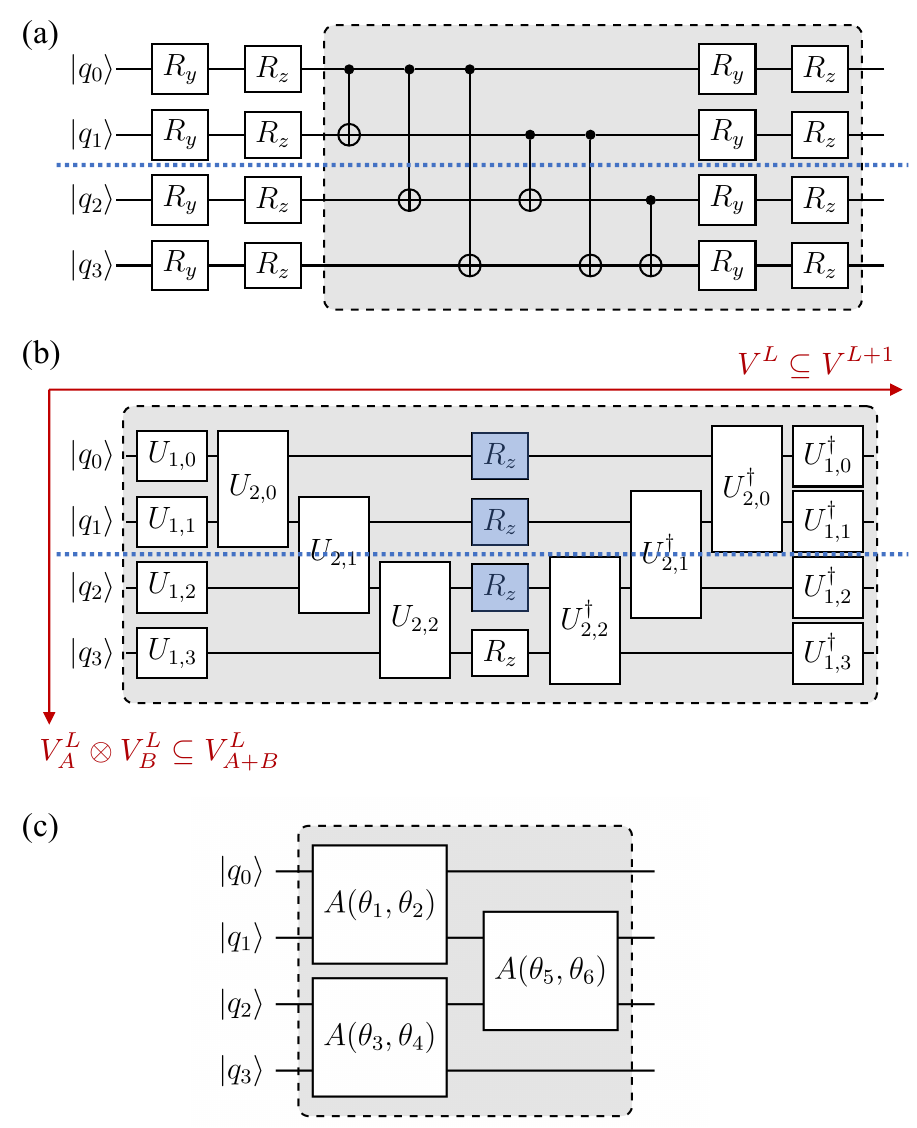}
\caption{Heuristically designed HEA and physics-constrained HEA. (a) $R_yR_z$ full (EfficientSU2) ansatz.
The $R_y$ full ansatz can be obtained by removing the columns of $R_z$ gates, while the $R_y$ linear ansatz can be derived from the $R_y$ full ansatz by further replacing the all-to-all CNOT gates by nearest neighbor CNOT gates. (b) XYZ1F (gates in white) and XYZ2F (gates in white and blue) ans\"{a}tze. The dashed box indicates a single repeating unit $U_l(\vec{\theta}_l)$, and all the parameters are omitted for brevity. The two inclusion constraints $V^L\subseteq V^{L+1}$ and $V^L_A\otimes V^L_B\subseteq V_{A+B}^L$ for extending HEA in different directions of quantum circuits are highlighted. The blue dashed line represents a partition
of the whole system into two subsystems $A$ and $B$. (c) An example of particle-number conserving ansatz introduced in Refs. \cite{barkoutsos2018quantum,gard2020efficient}.}\label{fig:ansatz}
\end{figure}

{\bf (3) Size-consistency}: since the exact wavefunction of a compound system $A+B$ consisting of two noninteracting subsystems $A$ and $B$ is multiplicatively separable, i.e. $|\Psi^{AB}\rangle=|\Psi^A\rangle|\Psi^B\rangle$, we require that $V_A^L\otimes V_B^L$ should be included in $V_{A+B}^{L}$ for any $L$, i.e. $V_A^L\otimes V_B^L \subseteq V_{A+B}^L$ (see Fig. \ref{fig:ansatz}b).
As the Hamiltonian of the composite system is $\hat{H}_{A+B}=\hat{H}_A+\hat{H}_{B}$,
the constraint ensures that the variational energy $E_{A+B}^{L}$ of the composite system will not be worse than the sum of the individually computed energies $E_{A}^L+E_{B}^L$, i.e. $E_{A+B}^L\le E_{A}^L + E_{B}^L$.
This size-consistency condition requires that for any 
$U_A(\vec{\theta}_{l,A})$ and $U_B(\vec{\theta}_{l,B})$, 
there exists a set of parameters $\vec{\theta}_{l,A+B}$ such that
$U_{A+B}(\vec{\theta}_{l,A+B})=U_A(\vec{\theta}_{l,A})\otimes U_B(\vec{\theta}_{l,B})$.

{\bf (4) Noninteracting limit}: in the limit that all the qubits are noninteracting, the eigenstates are given by product states. 
Thus, we require that for a good HEA,
$L=1$ should be sufficient for representing any product state, and
hence $L>1$ is only required for entangled states.

If we make an analogy between HEA for quantum wavefunction and neural networks (NN) for high-dimensional functions in classical computing, then the requirement (1) plays a similar role as the universal approximation theorem (UAT)\cite{cybenko1989approximation,hornik1989multilayer}
for NN. As shown in Fig. \ref{fig:ansatz}b, the two inclusion constraints (2) and (3) represent the constraints for extending HEA in two different directions of quantum circuits. To some extent, the systematic improvability is analogous to the ResNet\cite{he2016deep} in classical NN architecture, which 
uses NN to parameterize the residual and enables the use of very deep neural network in practice. 
In a similar spirit, we hope that an HEA with systematic improvability
can allow to use very deep quantum circuits. This turns out to be true
as demonstrated numerically in the later section.
The size-consistency constraint (3) introduced for HEA extends
the notion of size-consistency/size-extensivity in quantum chemistry\cite{pople1976theoretical,bartlett1981many,nooijen2005reflections} for the qualification 
and differentiation of many-body methods. A size-extensive method such as the coupled cluster theory\cite{bartlett1981many} can provide energies that grow linearly with the number of electrons in the system. This is mandatory for the application of a many-body method to large systems such as solids, because it guarantees that the quality of the energy will not deteriorate compared to that for small systems. This concept is therefore also essential for the scalability of the variational ansatz on quantum computers.

Some previously designed HEA are shown in Fig. \ref{fig:ansatz}a. 
The commonly used $R_y$ and $R_yR_z$ (EfficientSU2) ans\"{a}tze
with different entangling blocks\cite{kandala2017hardware} clearly fail to
meet these important requirements, in particular, constraints (2) and
(3). It is obvious that the $R_y$ ansatz can only represent real wavefunctions, whereas it is unclear whether the $R_yR_z$ ansatz is universal.
Recently, a 'cascade' ansatz is developed to satisfy the
condition $U_{l}(\vec{\theta}_{l})=I$ by adding the inverse
of the CNOT gates in $R_y$ ansatz into the repeating unit\cite{d2023challenges}.
But it fails to meet the constraints (3) and (4).
The HVA for model systems of the form $U_l(\vec{\theta}_l)=\prod_k e^{\ii\theta_{l,k}\hat{H}_k}$, where $\hat{H}_k$ is a component of the Hamiltonian of the system $\hat{H}=\sum_k \hat{H}_k$, satisfies constraints (2) and (3), but does not necessarily meet constraints (1) and (4), which are requirements for general-purpose HEA. Similarly, the separable-pair approximation (SPA) ansatz\cite{kottmann2022optimized,kottmann2023molecular} 
satisfies the size-consistency and is hardware-efficient, but not universal. 
Particle-number symmetry-preserving ans\"{a}tze have also been introduced. A typical example is the ASWAP ansatz\cite{barkoutsos2018quantum,gard2020efficient}
shown in Fig. \ref{fig:ansatz}c, where the following exchange-type two-qubit gate
$A(\theta,\phi)$ is used
\begin{eqnarray}
A(\theta,\phi) = 
\begin{bmatrix}
 1 & 0 & 0 & 0 \\
 0 & \mathrm{cos} \ \theta & e^{\ii\phi}\mathrm{sin} \ \theta & 0 \\
 0 & e^{-\ii\phi}\mathrm{sin} \ \theta & -\mathrm{cos} \ \theta  & 0 \\
 0 & 0 & 0 & 1
\end{bmatrix}.\label{eq:Agate}
\end{eqnarray}
Since only $|01\rangle$ and $|10\rangle$ are allowed to mix, this gate
preserve the particle number of the input state with well-defined
particle number. It is universal only within the Hilbert space with fixed number of electrons, but fails to satisfy the constraints (2) and (3), because the identity operator cannot be achieved by $A(\theta,\phi)$.
The ansatz using the following hop gate\cite{eddins2022doubling} 
\begin{eqnarray}
h(\varphi) = 
\begin{bmatrix}
 1 & 0 & 0 & 0 \\
 0 & \cos\varphi & -\sin\varphi  & 0 \\
 0 & \sin\varphi &  \cos\varphi  & 0 \\
 0 & 0 & 0 & -1
\end{bmatrix}\label{eq:hop}
\end{eqnarray}
also fails to satisfy the systematically improvability and size-consistency due to
the exactly same problem. In summary, to the best of our knowledge, an HEA satisfying all these constraints has not been proposed before. 
Our goal is to design an HEA satisfying these constraints and hence
it will not be particle number conserving in order to be universal,
which is advantageous in applications such as
computing Green's functions\cite{cai2020quantum,huang2022variational}.
For applications where the particle number is conserved,
such as computing the ground state of molecules,
we can apply penalties to enforce the correct particle numbers if necessary (vide post). Finally, we emphasize that apart from these theoretical constraints,
since most of existing quantum devices have very restricted qubit connectivity, an
additional important hardware constraint is that
the building block $U_l(\vec{\theta}_l)$
should be easily implemented on quantum 
devices with restricted connectivity.

\subsection{Physics-constrained HEA}
\begin{figure}[H]
\centering
\includegraphics[width=0.6\textwidth]{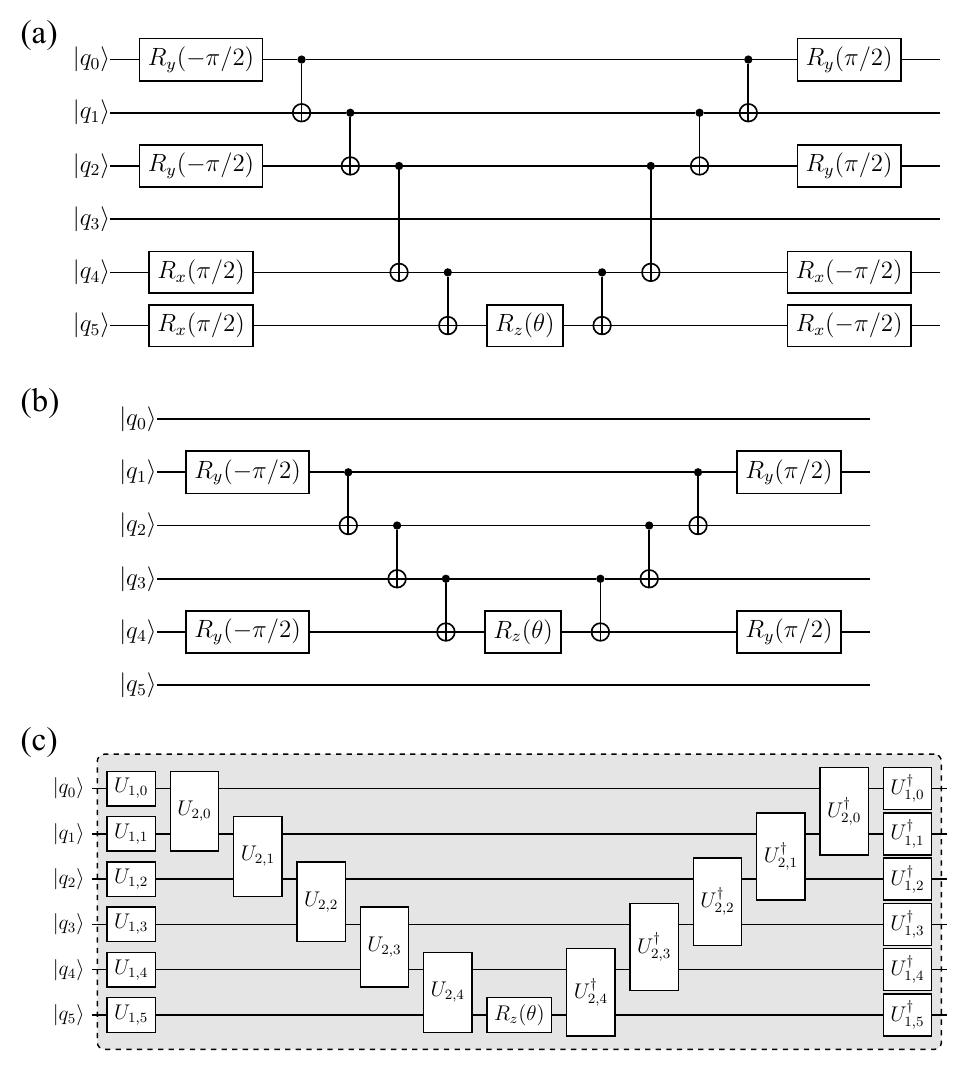}
\caption{Examples for Theorem 1 with six qubits. (a), (b) quantum circuits appeared in UCC
arising from the double excitation $a_0^\dagger a_2^\dagger a_4 a_5$
and the single excitation $a_1^\dagger a_4$, respectively;
(c) the XYZ1F circuit. Both (a) and (b) can be represented by (c) by appropriately choosing single-qubit gates $U_{1,k}$ from $\{I,R_x(\pi/2),R_y(-\pi/2)\mathrm{\; or\; }H\}$
and two-qubit gates $U_{2,k}$ from $\{I,\mathrm{CNOT},\mathrm{SWAP\; or\; iSWAP}\}$.
}\label{fig:theorem}
\end{figure}

The above requirements still leave a lot of degree of freedom in the design of HEA. Here we propose one possible HEA that satisfies these basic requirements, referred as physics-constrained HEA, and only requires linear qubit connectivity. It should be pointed out that such realization of physics-constrained HEA is not unique, and other realizations are certainly possible,
which is a subject of future investigations.
Our starting point is the wavefunction given 
by a product of exponential of Pauli operators
\begin{eqnarray}
|\Psi\rangle = e^{\ii\theta_L P_L}\cdots 
e^{\ii\theta_1 P_1}|\Phi_0\rangle
\triangleq
\prod_{l=1}^L e^{\ii \theta_l P_l}|\Phi_0\rangle,\label{eq:pauli}
\end{eqnarray}
which is the form for the UCC type ansatz.
A simple observation is that if one can choose all possible $P_l\in\{I,X,Y,Z\}^{\otimes N}$, this form of $U_l(\theta_l)$ (= $e^{\ii \theta_l P_l}$) is universal\cite{nielsen2010,evangelista2019exact}. It is also systematically improvable because
the choice $\theta_l=0$ gives an identity operator. However, it does not satisfy constraints
(3) and (4). Thus, if each layer of HEA has the ability to represent any $e^{\ii\theta_l P_l}$ with $P_l\in\{I,X,Y,Z\}^{\otimes N}$, the resulting ansatz will automatically satisfy constraints (1) and (2), and we only need to modify it to satisfy constraints (3) and (4).
The fact that any $e^{\ii\theta_l P_l}$ can be represented by a quantum circuit with CNOT 'staircases'\cite{nielsen2010} (see Figs. \ref{fig:theorem}a and \ref{fig:theorem}b) motivates us to design $U_l(\vec{\theta}_l)$ with a similar structure (see Fig. \ref{fig:theorem}c), where the forms of the single-qubit gates $U_1$ and two-qubit gates $U_2$ remain to be specified. We find the following sufficient condition for representing any $e^{\ii\theta_l P_l}$ by the circuit block in Fig. \ref{fig:theorem}c:
\begin{thm}
If $U_2$ include gates in $\{I,\mathrm{CNOT},\mathrm{SWAP\; or\; iSWAP}\}$, 
then the circuit block in Fig. \ref{fig:theorem}c can represent any $e^{\ii\theta_l P_l}$
with $P_l\in\{I,Z\}^{\otimes N}$. Furthermore, if
$U_1$ include gates in $\{I,R_x(\pi/2),R_y(-\pi/2)\mathrm{\; or\; }H\}$,
then the circuit block can represent any $e^{\ii\theta_l P_l}$
with $P_l\in\{I,X,Y,Z\}^{\otimes N}$.
\end{thm}
The proof of Theorem 1 is quite straightforward. We only give two concrete
examples in Fig. \ref{fig:theorem}, arsing from the double excitation $a_0^\dagger a_2^\dagger a_4 a_5$ and the single excitation $a_1^\dagger a_4$ in UCC, respectively.
It can be easily verified that $e^{\ii\theta Z_0Z_1Z_2Z_4Z_5}$ in Fig. \ref{fig:theorem}a is given by XYZ1F in Fig. \ref{fig:theorem}c with
$U_{2,k}=\mathrm{CNOT}$ for $k\in\{0,1,3,4\}$ and
$U_{2,2}=\mathrm{SWAP}/\mathrm{iSWAP}$, 
while $e^{\ii\theta Z_1Z_2Z_3Z_4}$ in Fig. \ref{fig:theorem}b
is given by $U_{2,0}=I$, $U_{2,k}=\mathrm{CNOT}$ for $k\in\{1,2,3\}$,
and $U_{2,4}=\mathrm{SWAP/\mathrm{iSWAP}}$. Other $e^{\ii \theta P_l}$ 
can be represented by a similar recipe.
The role of $\mathrm{SWAP}$ can be replaced by $\mathrm{iSWAP}$, which is easier to implement on some quantum computing platforms, such as superconducting
quantum devices\cite{kjaergaard2020superconducting}.

There are still infinitely many ways to parameterize $U_1$ and $U_2$ that satisfy this sufficient condition, since a general single-qubit (two-qubit) gate can be described by three (fifteen) parameters, and at most three CNOT gates are needed for general two-qubit gates\cite{vidal2004universal,vatan2004optimal}.
To minimize the number of parameters per circuit block and reduce
the number of native two-qubit gates in $U_2$, here we 
present a parameterization of $U_2$ with two parameters
\begin{eqnarray}
U_2(\theta,\phi) = [I\otimes R_y(\phi/2)]U_{\mathrm{fSim}}(\theta,\phi)
[I\otimes R_y(-\phi/2)],\label{eq:u2}
\end{eqnarray}
using the fSim gate $U_{\mathrm{fSim}}(\theta,\phi)$ native on some superconducting devices\cite{foxen2020demonstrating}
\begin{eqnarray}
U_{\mathrm{fSim}}(\theta,\phi)=
\begin{bmatrix}
1 & 0 & 0 & 0 \\
0 & \cos\theta & -\ii\sin\theta & 0 \\
0 & -\ii\sin\theta & \cos\theta & 0 \\
0 & 0 & 0 & e^{-\ii\phi} \\
\end{bmatrix},
\end{eqnarray}
which yields 
\begin{eqnarray}
U_2(0,0)=I,\quad U_2(-\pi/2,0)=\mathrm{iSWAP},\quad
U_2(0,\pi) = \mathrm{CNOT}.
\end{eqnarray}
A simple choice for the single-qubit gates $U_1$ to satisfy the sufficient condition is 
\begin{eqnarray}
U_1(\theta,\phi)=R_x(\theta)R_y(\phi).\label{eq:u1}
\end{eqnarray}
Eqs. \eqref{eq:u2} and \eqref{eq:u1} completely define a HEA, 
which can implement an exponential of any Pauli operator by appropriate choice of parameters.  
We will refer to it as XYZ1F in the following context, see Fig. \ref{fig:theorem}c, as an abbreviation for the combination of the three types of 
single-qubit rotation gates used and the two-qubit gates involving fSim gates.

It is easy to see that the XYZ1F ansatz, however, does not yet satisfy constraint (3) and (4). Fortunately, by simply replacing the single $R_z$ gate in the middle by a layer of $R_z$ gate, we can resolve this problem and obtain the final physics-constrained HEA, denoted by XYZ2F in Fig. \ref{fig:ansatz}b (with additional gates in blue). The size-consistency of XYZ2F can be seen as follows: Suppose the subsystems $A$ and $B$ contain the first two and the remaining two qubits, respectively,
then the wavefunction $|\Psi_A^L\rangle|\Psi_B^L\rangle$ formed by
a direct product of the two XYZ2F wavefunctions can be represented by an XYZ2F ansatz for the composite system with $U_{2,1}=I$ (see Fig. \ref{fig:ansatz}b).
Therefore, the ability of $U_2$ to become identity 
is essential from a size-consistent perspective, which
is missing in other HEA shown in Figs. \ref{fig:ansatz}a
and \ref{fig:ansatz}c. It can be verified that the constraint (4) is also satisfied by simply setting all $U_2$ to identity, such that for the $n$-th qubit the circuit block gives a universal single-qubit gate\cite{nielsen2010}
\begin{eqnarray}
U = R_x(\theta_{0})R_y(\theta_{1})R_z(\theta_{2})
R_y^\dagger(\theta_{1})R_x^\dagger(\theta_{0}).
\end{eqnarray}
Another advantage of the size-consistent modification is that terms such 
as $e^{\ii (\theta_1 X_0X_1 + \theta_2 X_2Z_3Z_4 + \theta_3 Y_5)}$
can be implemented by a single layer in XYZ2F. In contrast, $e^{\ii \theta_1 X_0X_1}e^{\ii \theta_2 X_2Z_3Z_4}e^{\ii\theta_3 Y_5}$ needs to be implemented by three consecutive blocks
in XYZ1F. This will greatly reduce the number of layers required
to represent certain states, as will be shown numerically for the ground
state of the Heisenberg model and some typical molecules. 

In summary, the constructed XYZ2F ansatz satisfies all the four
fundamental constraints. The number of parameters in one layer of XYZ2F is $(5N-2)$,  where $3N$ and $2(N-1)$ are for single-qubit and two-qubit gates, respectively. Comparing the exponential $e^{\ii \theta_l P_l}$ with $P_l\in\{I,X,Y,Z\}^{\otimes N}$, it is seen that all the $4^N$ discrete choices of $P_l$ are now embedded into a continuous space of operators $U_l(\vec{\theta}_l)$ specified by $O(N)$ parameters. 
Therefore, it can be viewed as an adaptive ansatz,
where the operator pool contains all $4^N$ Pauli operators  
rather than given by UCCSD as in other adaptive methods\cite{ryabinkin2018qubit,grimsley2019adaptive,tang2021qubit,yordanov2021qubit}. In Table \ref{tab:hea_cost}, we compare
different HEA in terms of
the numbers of parameters $N_{\mathrm{param}}$, 
two-qubit gates $N_2$,
single-qubit gates $N_1$,
and the circuit depth $D$
as a function of the layer $L$ and the number of qubits $N$.
The number of parameters in all of them scales as $O(NL)$.
One disadvantage of XYZ1F and XYZ2F is that the circuit depths of
XYZ1F and XYZ2F is larger than other HEA due to the
use of the staircase structure.
In principle, other low-depth architectures such as the brickwall structure 
can be used in the construction of physics-constrained HEA.
We are exploring such possibility and the results will be reported elsewhere.

\begin{table}[H]
\centering
\caption{Comparison of different HEA in terms of
the numbers of parameters $N_{\mathrm{param}}$, 
two-qubit gates $N_2$,
single-qubit gates $N_1$,
and the circuit depth $D$
as a function of the layer $L$ and the number of qubits $N$.
When counting $N_2$ and $D$, we assume the ASWAP and fSim gates are native gates.
}\label{tab:hea_cost}
\begin{tabular}{ccccc}
        \hline\hline
        ansatz       &  $N_{\mathrm{param}}$ & $N_2$ & $N_1$ & $D$ ($N\geq3,L\geq1$)  \\ 
        \hline 
        $R_y$ linear      &  $N(L+1)$      &  $(N-1)L$    &  $N(L+1)$      & $N+3L-2$        \\ 
        $R_y$ full        &  $N(L+1)$      &  $N(N-1)L/2$ &  $N(L+1)$      & $NL+N+L-2$      \\ 
        $R_yR_z$ full     &  $2N(L+1)$     &  $N(N-1)L/2$ &  $2N(L+1)$     & $NL+N+2L-2$     \\ 
        ASWAP             &  $2(N-1)L$     &  $(N-1)L$    &   0            & $2L$            \\ 
        XYZ1F             &  $(4N-1)L$     &  $2(N-1)L$   &  $(8N-3)L$     & $(4N+3)L$       \\ 
        XYZ2F             &  $(5N-2)L$     &  $2(N-1)L$   &  $(9N-4)L$     & $(4N+3)L$       \\ 
\hline\hline
\end{tabular} 
\end{table}

\subsection{Layerwise optimization algorithm}
To take full advantage of the systematic improvability of XYZ1F and XYZ2F,
the parameters in HEA are optimized in a layerwise way using the Algorithm \ref{algorithm}. 
This is different from optimization with random initialization for all parameters, which has been shown to easily suffer from the problems of low-quality local minima and barren plateaus\cite{mcclean2018barren}. Our algorithm has two key features. First, we retain the optimized parameters from the previous step as the initial guess for the $L-1$ layers in the current step. Second, we generate $n$ (about 10) sets of random parameters with different step sizes for the $L$-th layer. Each set of parameters $\vec{\theta}_L$ is obtained
from $\vec{\theta}_L=\frac{\vec{u}}{\max_i|u_i|}\delta$, where $\vec{u}$ are random numbers in $[-1,1]$ and $\delta$ is a predefined
step size $\delta\in\{\frac{2\pi}{2^k}\}_{k=0}^{5}\cup \{0\}$.
This guarantees that the optimized energy decreases monotonically 
with respect to $L$ for XYZ1F and XYZ2F. For consistency, the layerwise optimization method is also employed in calculations using other HEA.
We implement all the HEA using MindQuantum\cite{mq_2021}.
Numerical optimization in VQE employed the Broyden–Fletcher–Goldfarb–Shanno (BFGS) method implemented in Scipy\cite{virtanen2020scipy}. 
As shown in the Supplementary Materials, during the the optimization,
the decrease of the energy is fast at the beginning of iterations 
and then slows down. Thus, we set the maximal number of iterations to be 3000 for a given layer. 

\begin{algorithm}[H]
\caption{Layerwise optimization strategy}
\begin{algorithmic}
\Require $L_{\max}$, $|\Phi_{0}\rangle$;        
\State Initialization: $L=0,|\Psi\rangle=|\Phi_{0}\rangle$;
\While {$L \le L_{\max}$}
    \State Set $L \mathrel{+}= 1$;
    \State Generate $n$ sets of random parameters $\{\vec{\theta}_{L}^{(i)}\}_{i=1}^n$ with different step sizes for the $L$-th layer;
    \State Combine $\{\vec{\theta}_{1}, \dots, \vec{\theta}_{L-1}\}$ obtained in the previous step as initial guess for the $L-1$ layers;
    \State Perform $n$ independent VQE with $|\Psi\rangle=U_L\cdots U_1|\Phi_0\rangle$ in parallel;
    \State Collect $E_{L}=\min_{i} E^{(i)}_{L}$ and optimized $\{\vec{\theta}_{1}, \dots, \vec{\theta}_{L}\}$;
\EndWhile
\Ensure $\{E_{1},\cdots,E_{L_{\max}}\}$ and $|\Psi\rangle = U_{L_{\max}}\cdots U_{1}|\Phi_{0}\rangle$.
\end{algorithmic}\label{algorithm}
\end{algorithm}

\begin{figure}[H]
\centering
 	\includegraphics[width=0.9\textwidth]{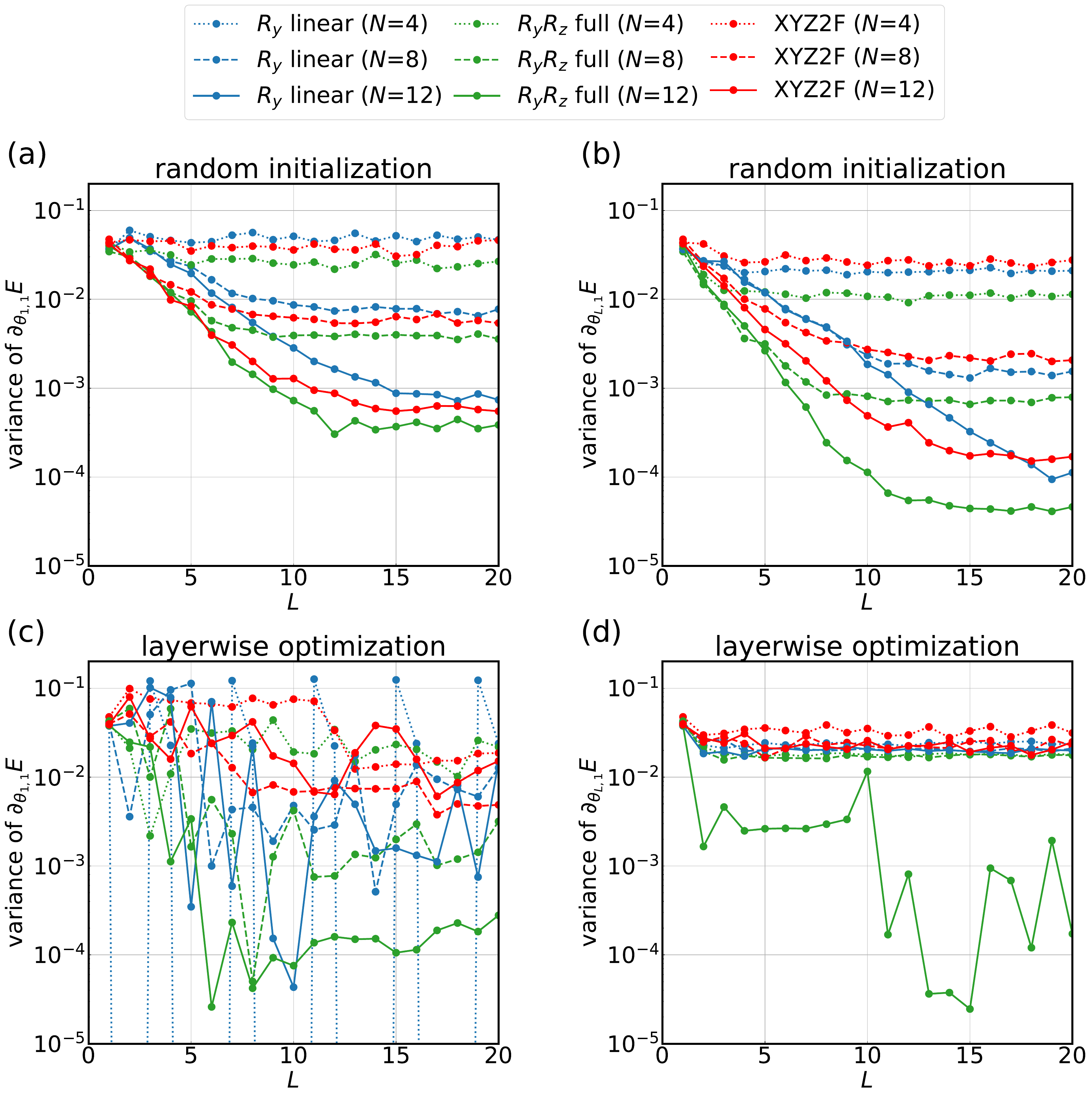} 
	\caption{Variance of energy gradients for the one-dimensional Heisenberg model.
(a) and (b): the variances of energy gradients for the first parameter in the first layer $\theta_{1,1}$ and that in the $L$-th layer $\theta_{L,1}$ are computed with random initialization of all parameters for 500 times.
For all the HEA, an exponential vanishing of gradients with respect to the number of qubits $N$ and the number of layers $L$ are observed, consistent with the conclusions in Ref. \cite{mcclean2018barren}.
(c) and (d): instead of random initialization, the parameters for the 
previous $L-1$ layers are the optimized parameters, while those for the $L$-th layer are initialized randomly for 500 times. Compared with random initialization, the layerwise optimization alleviates the barren plateau problem.
}\label{fig:barren_plateau}
\end{figure}

Figure \ref{fig:barren_plateau} displays the variance of energy gradients for
the one-dimensional Heisenberg model (see the next section). We observe
the exponential vanishing of gradients with respect to the number of qubits $N$ and the number of layers $L$ with random initialization, consistent with the conclusions in Ref. \cite{mcclean2018barren}. In contrast, the layerwise optimization alleviates the barren plateau problem in particular for XYZ2F.

\section{Results}
\subsection{Heisenberg model}
We first use the one-dimensional Heisenberg model with
open boundary condition, whose Hamiltonian is $\hat{H}=-\frac{1}{2}J\sum_{n=1}^{N-1}\vec{\sigma}_{n}\cdot\vec{\sigma}_{n+1}$, to study the effectiveness of the constructed HEA. 
In Table  \ref{tab:consistency}, we
perform a size-consistent
test\cite{pople1976theoretical,nooijen2005reflections} for different HEA by applying them to a composite Heisenberg model (denoted by 6+6) consisting of two noninteracting subsystems with six sites. 
The parameters in the wavefunction of
the composite system are taken from the optimized parameters for the subsystem. 
If an ansatz is size-consistent, then the ground-state energy per site $e_{N}^L$
should be the same for the whole system and the subsystem, i.e. $e_{6+6}^{L}=e_{6}^L$ for any $L$. Notably, only XYZ2F satisfies this condition, while other HEA violate it significantly. The additional entangling gates between subsystems in other HEA severely degrade the quality of the approximation in the total system, as can be seen from the significant increase of infidelity
($1-F_N^L$ with $F_{N}^L=|\langle\Psi_{N}^L|\Psi_{N}^*\rangle|^2$,
where $|\Psi_{N}^*\rangle$ is the exact wavefunction) in Table \ref{tab:consistency}. In particular, the additional CNOT gates in the $R_y$ full ansatz (see Fig. \ref{fig:ansatz}a) makes the fidelity between the approximate state and the ground state almost vanish. On the contrary, the infidelity for XYZ2F is well-controlled, that is, if the fidelity $F_6^L$ is $1-\epsilon$, where $\epsilon$ 
is a small number (0.00085 for $L=2$ in Table \ref{tab:consistency}),
then the fidelity for the total system is $(1-\epsilon)^2$,
and thus the infidelity $1-F_{6+6}^L$ is about $2\epsilon$.
Therefore, an interesting topic for future studies is 
to use parameters optimized from small systems 
as an initial guess of XYZ2F for large systems.

\begin{table}[H]
    \caption{Size-consistency test: ground-state energy per site $e_{N}^L$ and infidelities ($1-F_N^L$ with $F_{N}^L=|\langle\Psi_{N}^L|\Psi_{N}^*\rangle|^2$)
    obtained by different HEA  with $L=2$ and $L=4$
    for a composite Heisenberg model (denoted by 6+6), which consists of two noninteracting subsystems with six sites. The parameters of the total system are taken from the optimized parameters of the subsystems.
    Errors with respect to the exact $e_{N}^*$ obtained from exact diagonalization are shown
    in parenthesis. Only XYZ2F satisfies $e_{6+6}^{L}=e_{6}^L$, while
    other HEA are not size-consistent.}\label{tab:consistency}
    \centering
    \scalebox{1.0}{
    \begin{tabular}{cccccc}
        \hline \hline
        system            & $R_{y}$ linear & $R_{y}$ full & $R_yR_z$ full & ASWAP & XYZ2F \\ \hline
        $e_6^{L=2}$       & -0.78333       & -0.78333     &  -0.80606     & -0.80273  &  -0.83054        \\
                          & (0.04786)      & (0.04786)    & (0.02514)     & (0.02846) &  ({\bf 0.00065}) \\
        $e_{6+6}^{L=2}$   & -0.52002       & -0.51368     &  -0.64791     & -0.52267  &  -0.83054        \\  
                          & (0.31118)      & (0.31751)    & (0.18328)     & (0.30853) & ({\bf 0.00065})  \\
        $e_6^{L=4}$       & -0.82988       & -0.82988     & -0.83089      & -0.83119  & -0.83119         \\
                          & (0.00132)      & (0.00132)    & (0.00030)     & (0.00000) & ({\bf 0.00000})  \\
        $e_{6+6}^{L=4}$   & -0.33247       & -0.35020     & -0.53154      & -0.63073  & -0.83119         \\
                          & (0.49873)      & (0.48099)    & (0.29966)     & (0.20047) & ({\bf 0.00000})  \\ \hline
        $1-F_6^{L=2}$     & 0.17623        & 0.17623      & 0.03743       & 0.05260   & {\bf 0.00085}    \\ 
        $1-F_{6+6}^{L=2}$ & 0.77773        & 1.00000      & 1.00000       & 0.95358   & {\bf 0.00170}    \\
        $1-F_{6}^{L=4}$   & 0.00205        &  0.00205     & 0.00037       & 0.00000   & {\bf 0.00000}    \\
        $1-F_{6+6}^{L=4}$ & 0.93368        &  1.00000     & 0.93196       & 0.79871   & {\bf 0.00000}    \\ 
        \hline\hline    
    \end{tabular}}
\end{table}

\begin{figure}[H]
\centering
 	\includegraphics[width=1.0\textwidth]{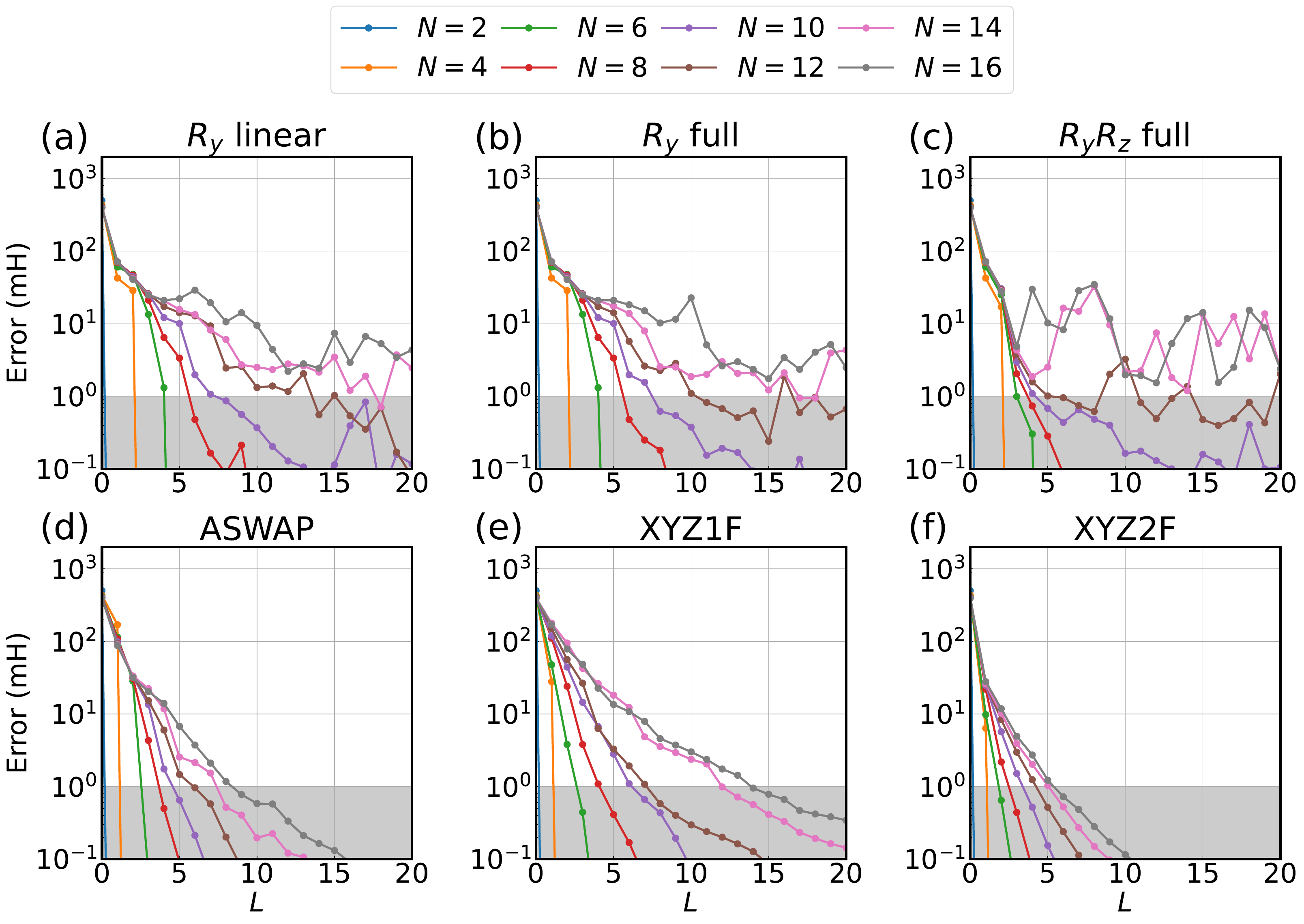} 
	\caption{Convergence of the ground-state energy per site $e_N^L$ obtained by different HEA ($R_y$ linear, $R_y$ full, $R_yR_z$ full, ASWAP, XYZ1F and XYZ2F) as a function of the number of layers $L$ for one-dimensional antiferromagnetic Heisenberg models ($J=-1$) with different number of sites $N$. The exact values $e_N^*$ are obtained from exact diagonalizations. While both XYZ1F and XYZ2F are systematically improvable, the size-consistency of XYZ2F makes it superior to XYZ1F. Other HEA do not converge monotonically, and become increasingly difficult to achieve a target accuracy as the system size increases (except for ASWAP). The shaded regions represent the region within the chemical accuracy (1 mH).}\label{fig:convergence}
\end{figure}

Figure \ref{fig:convergence} shows the 
convergence of the ground-state energy per site $e_N^L$ obtained
by different HEA for antiferromagnetic Heisenberg models
as a function of the number of layers $L$ starting from
a N\'{e}el state as reference $|\Phi_0\rangle$. We find that
both XYZ1F and XYZ2F are systematically improvable as expected,
and the effect of the size-consistent modification in XYZ2F is dramatic, which significantly reduces the number of layers needed to reach certain accuracy. In contrast,
other HEA do not converge monotonically and become increasingly difficult to 
converge as the system size $N$ increases (except for ASWAP). As shown in Fig. \ref{fig:convergence}, the oscillatory behavior reveals a severe problem of these HEA in practical applications, that is, even if they have reached a certain accuracy with $L$ layers, the accuracy with $L+1$ layers may be worse. The comparison with XYZ1F/XYZ2F suggests that it is the violation of constraints (2) and (3) that causes these heuristically designed HEA to perform poorly as $N$ increases.

\begin{figure}[H]
\centering
 	\includegraphics[width=0.9\textwidth]{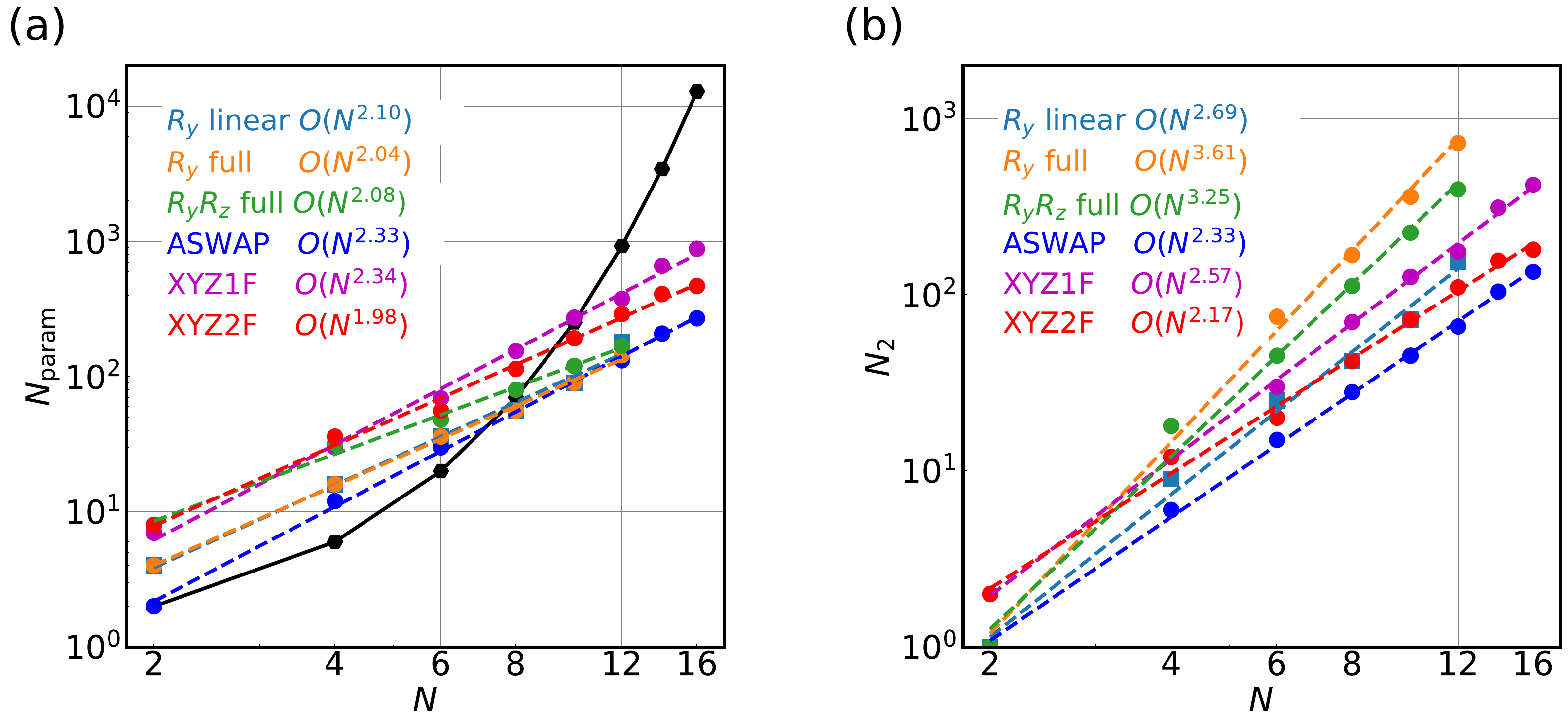} 
	\caption{The number of variational parameters $N_{\mathrm{param}}$ (a) and the number of two-qubit gates $N_2$ (b) required to reach 1 milli-Hartree as a function of the number of sites $N$ for one-dimensional antiferromagnetic Heisenberg models ($J=-1$). The dimension of the Hilbert space $\binom{N}{N/2}$ for different $N$ is shown for comparison. The fitted scaling $O(N^b)$ is shown by dashed lines.}\label{fig:scaling}
\end{figure}

In Fig. \ref{fig:scaling}a, we display the number of variational parameters $N_{\mathrm{param}}$ to reach 1 milli-Hartree as a function of the number of sites $N$, compared against the dimension of the Hilbert space $\binom{N}{N/2}$ for different $N$. It is seen that for this system, the number of variational parameters to reach 1 milli-Hartree scales as $O(N^{1.98})$ for XYZ2F. The scaling for XYZ1F and ASWAP are quite similar, but
the ASWAP has a much smaller prefactor. Figure \ref{fig:scaling}b
shows the scaling of the number of two-qubit gates $N_2$ required to reach 1 milli-Hatree for different $N$. It is seen that XYZ2F has the lowest scaling $O(N^{2.17})$, albeit with a larger prefactor than ASWAP. Therefore, future optimization of the
layout may further lead to more economic physics-constrained HEA.

\subsection{Molecules}
Next we examine the performance of the constructed HEA for realistic systems.
For molecules, the molecular integrals were generated using the PySCF package\cite{sun2018pyscf} in a minimal STO-3G basis. The Jordan-Wigner fermion-to-qubit transformation\cite{jordan1928pauli} was carried out using OpenFermion\cite{mcclean2020openfermion}, where occupation number vectors (ONVs) for fermions $|n_{0}n_{\bar{0}}n_{1}n_{\bar{1}}\cdots n_{k}n_{\bar{k}}\rangle$ ($n\in \{0,1\}$ and the symbol $\bar{i}$ represents $i$-th $\beta$ spin-orbital) is mapped to ONVs for qubits $|q_{0}q_{1}q_{2}q_{3}\cdots q_{2k-1}q_{2k}\rangle$. 

Table \ref{tab:consistency_h4} shows the size-consistency test for a
composite hydrogen ladder composed of two \ce{H4} chain ($R_{\mathrm{H-H}}=$1.5 \AA) separated by a distance of 100 \AA. It is well-known that 
classical spin-restricted
coupled cluster singles and doubles (CCSD) is size-consistent/extensive in
this case. However, it is seen that the commonly used 
Trotterized unitary CCSD (UCCSD)
using the canonical molecular orbitals (CMO) has a small size-consistency
error. Here, we only consider a single Trotter step. The Trotterization
makes UCCSD lose the property of orbital invariance. Since the CMO
of the \ce{H4} ladder, which is delocalized among all hydrogen, 
is different from that of the \ce{H4} monomer, the UCCSD wavefunction
cannot be exactly factorized into a product of two wavefunctions.
The Trotterized UCCSD can become size-consistent only with 
localized molecular orbitals (LMO) with a proper ordering of qubits.
Similarly, we can only expect size-consistency for HEA using localized
orbitals. Table \ref{tab:consistency_h4} shown the results obtained with $L=6$ using orthonormalized atomic orbitals (OAO). It is clear that only
XYZ2F is size-consistent in this case.

\begin{table}[H]
    \caption{Size-consistency test for a
composite system composed of two \ce{H4} chains ($R_{\mathrm{H-H}}=$1.5 \AA) separated by a distance of 100 \AA. The ground-state energies
for the monomer and the entire system are obtained by
classical coupled cluster singles and doubles (CCSD)
and single-step Trotterized unitary CCSD (UCCSD) using
canonical molecular orbitals (CMO). For HEA, 
the results are obtained with $L=6$ using 
orthonormalized atomic orbitals (OAO).
The optimized parameters for \ce{H4} are taken
as the parameters for the \ce{H4}-\ce{H4} ladder.
Errors with respect to the FCI are shown in parenthesis.}\label{tab:consistency_h4}
    \centering
    \scalebox{0.85}{
    \begin{tabular}{cccccccc}
        \hline \hline
        \multirow{2}*{system}                           &CCSD       & UCCSD    & $R_{y}$ linear & $R_{y}$ full & $R_yR_z$ full & ASWAP  & XYZ2F    \\ 
                                         & (CMO)      & (CMO)     & (OAO)          & (OAO)        & (OAO)         & (OAO)  & (OAO)    \\ \hline
        \multirow{2}*{$E_{\mathrm{H_{4}}}$}       & -1.99762   & -1.99460  &  -1.97672      &  -1.99045   & -0.92429     & -1.99113   & -1.99560  \\
                                         & (-0.00147) & (0.00155) &  (0.01943)     &  (0.00570)  & (1.07186)    & (0.00502)  & \bf{(0.00055)}  \\
        \multirow{2}*{$E_{\mathrm{H_{4}\textnormal{-}H_{4}}}$} & -3.99525   & -3.98918  & -2.52810       & -2.36808   & -1.85371     & -3.48508   & -3.99119  \\  
                                         & (-0.00295) & (0.00312) &  (1.46420)     & (1.62422)  & (2.13859)    &  (0.50722) & \bf{(0.00111)}  \\
        $E_{\mathrm{H_{4}\textnormal{-}H_{4}}}- 2E_{\mathrm{H_{4}}}$
                                         & 0          & 2.3$\times10^{-5}$  & 1.43          & 1.61      & -5.1$\times 10^{-3}$    & $5.0\times 10^{-1}$ & 0 \\
        \hline\hline    
    \end{tabular}}
\end{table}

\begin{figure}[H]
\centering
 	\includegraphics[width=0.9\textwidth]{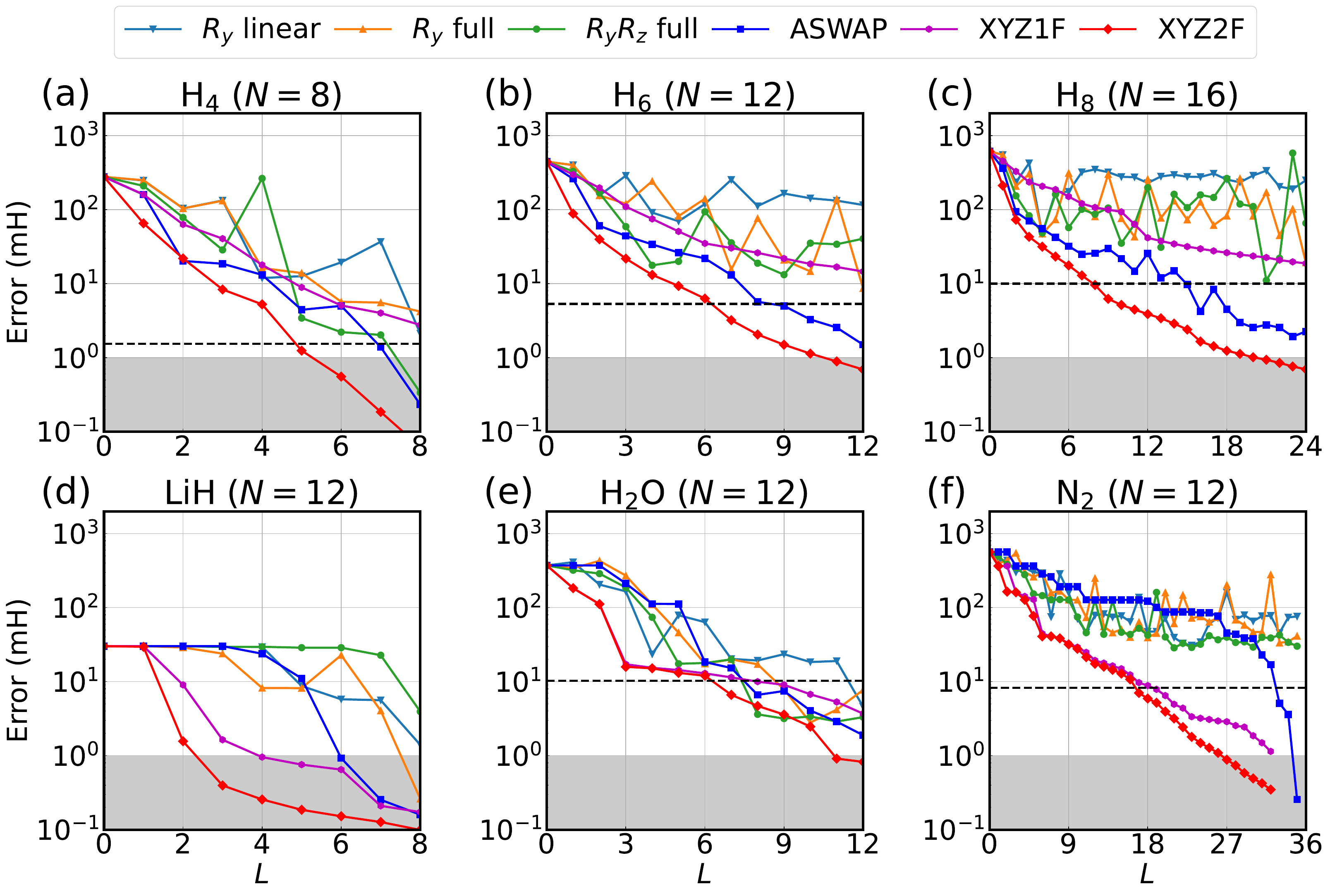} 
	\caption{Ground-state energy convergence with respect to exact diagonalization results
 as a function of the number of layers $L$ for molecules: (a)-(c) hydrogen chains (H$_4$, H$_6$, and H$_8$) with interatomic distance $R_{\mathrm{H\textnormal{-}H}}=1.5$ {\AA}; (d) LiH ($R_{\mathrm{Li\textnormal{-}H}}=2.0$ {\AA}); (e) H$_2$O ($R_{\mathrm{O\textnormal{-}H}}=2.0$ {\AA} and $\theta_{\mathrm{H\textnormal{-}O\textnormal{-}H}}=104.5$ degree); (f) N$_2$ ($R_{\mathrm{N\textnormal{-}N}}=2.0$ {\AA}). For \ce{H2O}, the O1s orbital is frozen, while the N1s and N2s orbitals are frozen for \ce{N2}. The black dashed lines represent UCCSD results, while the shaded regions represent the region within the chemical accuracy (1 mH).}\label{fig:scalability}
\end{figure}

Figure \ref{fig:scalability} shows the comparison of different HEA for the convergence of ground-state energies as a function of the number of layers
$L$ for hydrogen chains (\ce{H4}, \ce{H6}, and \ce{H8}), 
LiH, \ce{H2O}, and \ce{N2}, which were commonly used to benchmark the performance of quantum computing techniques\cite{kandala2017hardware,lee2018generalized,
d2023challenges,arute2020hartree,
magoulas2023linear}. For the stretched hydrogen chains, the OAO were used 
due to the faster convergence and the reference state was a N\'{e}el state, e.g. $|\Phi_{0}\rangle = |10011001\rangle$ for \ce{H4}. 
As the number of hydrogen increases, the number of Slater determinants with large coefficients in the expansion of the exact ground state increases significantly. 
As shown in Fig. \ref{fig:scalability}, only XYZ2F converges monotonically to chemical accuracy, while other HEA perform poorly for \ce{H6} and \ce{H8}. The number of layers needed to achieve chemical accuracy increases roughly linearly with the system size for XYZ2F. 

For LiH, \ce{H2O}, and \ce{N2}, all simulated with 12 qubits using the restricted Hartree-Fock (RHF) orbitals, the convergence behavior is quite different, reflecting the very different electronic structures. For LiH at $R_{\mathrm{Li\textnormal{-}H}}=2.0$ {\AA}, the exact ground state is dominated by the Hartree-Fock configuration (about 95\%), and thus XYZ2F quickly reaches chemical accuracy with $L=3$. For \ce{H2O} and \ce{N2} at stretched geometries upon dissociation, which are typical examples of strong electron correlations in quantum chemistry, the convergence is slower. In the calculation of \ce{H2O},
we added a penalty term for preserving the particle number, viz. $\hat{H}' = \hat{H}+\beta(\hat{N}_{\uparrow}-n_{\uparrow})^{2}+\beta(\hat{N}_{\downarrow}-n_{\downarrow})^{2}$ with $\beta=1.0$. In the calculations of \ce{N2}, we further added a penalty for the total spin, viz. $\hat{H}' = \hat{H}+\beta(\hat{N}_{\uparrow}-n_{\uparrow})^{2}+\beta(\hat{N}_{\downarrow}-n_{\downarrow})^{2}+\beta\hat{S}_{+}\hat{S}_{-}$ with $\beta=1.0$. For the ASWAP ansatz,
a larger $\beta=3.0$ is used, otherwise it will not converge to the ground
state. As shown in Fig. \ref{fig:scalability}, the number of layers required for XYZ2F to reach chemical accuracy is 11 and 27, respectively.
For the most challenging molecule \ce{N2}, we find that other HEA are difficult to converge to chemical accuracy.

\begin{figure}[H]
\centering
 	\includegraphics[width=0.7\textwidth]{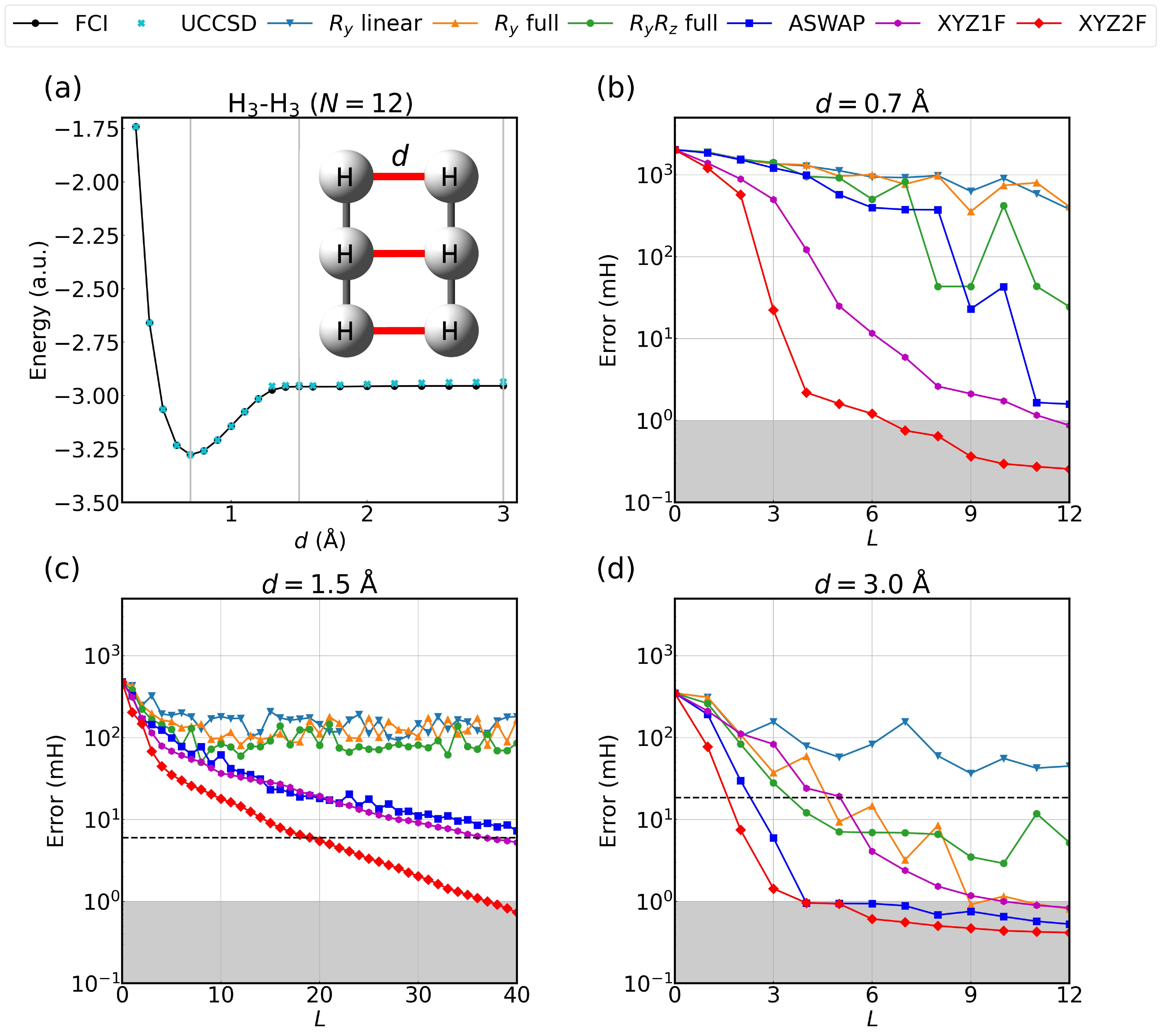} 
	\caption{A two-dimensional ladder \ce{H3}-\ce{H3} where $R_{\mathrm{H\textnormal{-}H}}=1.5$ \AA ~within each monomer. (a) Potential energy curve described by full configuration interaction (FCI) and unitary CCSD (UCCSD). Note that the performance of UCCSD deteriorates as $d$ exceeds 1.3 \AA. (b)-(d) Convergence of different HEA for the ground-state energy at three representative distances (0.7 \AA,  1.5 \AA, and 3.0 \AA). The black dashed lines represent UCCSD results. The error of UCCSD at $d=0.7$ \AA ~is below 1 micro-Hartree.}\label{fig:pes}
\end{figure}

Finally, we test the performance of HEA on a more challenging two-dimensional system, a \ce{H3}-\ce{H3} ladder (see Fig. \ref{fig:pes}). We should mention
that $R_y$ linear, ASWAP, XYZ1F, and XYZ2F are designed to be one-dimensional,
and we are working on the two-dimensional extensions. But we can examine their
performance on a two-dimensional system. The convergence of different HEA for the ground-state energy at three representative distances (0.7 \AA,  1.5 \AA, and 3.0 \AA) are shown in Fig. \ref{fig:pes}. Both in the equilibrium ($d=0.7$ \AA) and in
the dissociation region ($d=3.0$ \AA), 
XYZ2F converge to the chemical accuracy easily,
because the systems can be viewed as close to three hydrogen molecules
and two \ce{H3}, respectively. At $d=$1.5 \AA, which equals to
$R_{\mathrm{H\textnormal{-}H}}$ within the monomer, the system is most difficult.
We find that XYZ2F require about 37 layers to converge to the chemical accuracy,
while XYZ1F and ASWAP converge much more slowly. Other HEA completely fail.
It is seen that ensuring the universality, systematical improvability,
and size-consistency is important for the good performance of HEA
even in this challenging case. Therefore, we expect that
by extending the physics-constrained HEA to two-dimensional systems,
better performance can be obtained.

\section{Conclusion}
In this work, we introduced a new way to design HEA 
by satisfying fundamental constraints, inspired by the physics-constrained way to design non-empirical XC functionals in DFT\cite{kaplan2023predictive}.
The developed physics-constrained HEA - XYZ2F is superior to other heuristically designed
HEA in terms of both accuracy and scalability. In particular, numerical tests show the promise of XYZ2F for challenging realistic molecules with strong electron correlation. The better scalability of XYZ2F is attributed to the satisfaction of the systematic improvability and size-consistency. 
Our results suggest that incorporating physical constraints
into the design of HEA is a promising path towards
designing efficient variational ans\"{a}tze for solving many-body problems on quantum computers. 

One disadvantage of XYZ2F is its high circuit depth, due to the use of
the staircase structure. This stems from the requirement that
the circuit block can represent any exponential of a Pauli operator,
which is a sufficient condition for the universality. However,
this is not a necessary condition, other conditions for the universality
can be imposed, which may lead to lower circuit depth.
Another very interesting direction is that while we only consider
one-dimensional HEA in this work, the concepts of
physics-constrained HEA can be extended to construct
HEA for higher dimensions. We are exploring these directions.
It is conceivable that this work will inspire other realizations of HEA that satisfy these basic constraints, probably with more interesting properties such
as less parameters, faster convergence, better trainability, and more versatile qubit connectivity.

\begin{acknowledgement}

The authors acknowledge helpful comments by Jakob Kottmann and Mario Motta.
This work was supported by the National Natural Science Foundation of China (Grants No. 21973003) and the Fundamental Research Funds for the Central Universities.

\end{acknowledgement}

\begin{suppinfo}
%
%
%

Additional information including the energy convergence behavior of XYZ2F, the convergence with respect to the number of parameters, two-qubit gates count, and circuit depth, comparison of
the use of different orbitals and references, and results obtained
from noisy simulations.

This information is available free of charge via the Internet at http://pubs.acs.org. 

\end{suppinfo}

\bibliography{reference}

\end{document}